\documentclass[fleqn,usenatbib]{mnras}

\usepackage{newtxtext,newtxmath}

\usepackage[T1]{fontenc}

\DeclareRobustCommand{\VAN}[3]{#2}
\let\VANthebibliography\thebibliography
\def\thebibliography{\DeclareRobustCommand{\VAN}[3]{##3}\VANthebibliography}

\usepackage{graphicx}	% Including figure files
\usepackage{amsmath}	% Advanced maths commands
\usepackage{amssymb}	% Extra maths symbols
\usepackage{multirow}
\title[Gamma-Ray Burst 200826A]{An early peak in the radio light curve of short-duration Gamma-Ray Burst 200826A }

\author[L. Rhodes et al.]{
Lauren Rhodes,$^{1,2}$\thanks{E-mail: lauren.rhodes@physics.ox.ac.uk}
Rob Fender,$^{1,3}$
David R.A. Williams,$^{4}$
Kunal Mooley $^{5,6}$
\\
$^{1}$ Astrophysics, Department of Physics, University of Oxford, Keble Road, Oxford OX1 3RH, UK\\
$^{2}$ Max-Planck-Institut f\"{u}r Radioastronomie, Auf dem H\"{u}gel 69, 53121 Bonn, Germany\\
$^{3}$ Department of Astronomy, University of Cape Town, Private Bag X3, Rondebosch 7701, South Africa\\
$^{4}$ Jodrell Bank Centre for Astrophysics, School of Physics and Astronomy, The University of Manchester, Manchester, M13 9PL, UK\\
$^{5}$ National Radio Astronomy Observatory, Socorro, New Mexico 87801, USA\\
$^{6}$ Cahill Center for Astronomy, MC 249-17, California Institute of Technology, Pasadena, CA 91125, USA
}

\date{Accepted XXX. Received YYY; in original form ZZZ}

\pubyear{2015}

\begin{document}
\label{firstpage}
\pagerange{\pageref{firstpage}--\pageref{lastpage}}
\maketitle

\begin{abstract}
We present the results of radio observations from the eMERLIN telescope combined with X-ray data from \textit{Swift} for the short-duration Gamma-ray burst (GRB) 200826A, located at a redshift of 0.71. The radio light curve shows evidence of a sharp rise, a peak around 4--5 days post-burst, followed by a relatively steep decline. We provide two possible interpretations based on the time at which the light curve reached its peak. (1) If the light curve peaks earlier, the peak is produced by the synchrotron self-absorption frequency moving through the radio band, resulting from the forward shock propagating into a wind medium and (2) if the light curve peaks later, the turn over in the light curve is caused by a jet break. In the former case we find a minimum equipartition energy of $\sim3\times10^{47}$ erg and bulk Lorentz factor of $\sim5$, while in the latter case we estimate the jet opening angle of $\sim9-16\degr$. Due to the lack of data, it is impossible to determine which is the correct interpretation, however due to its relative simplicity and consistency with other multi-wavelength observations which hint at the possibility that GRB 200826A is in fact a long GRB, we prefer scenario one over scenario two. 
\end{abstract}

%%%%%%%%%%%%%%%%%%%%%%%%%%%%%%%%%%%%%%%%%%%%%%%%%%

%%%%%%%%%%%%%%%%% BODY OF PAPER %%%%%%%%%%%%%%%%%%

\section{Introduction}
Gamma-Ray Bursts (GRBs) are flashes of gamma-rays that are thought to be produced from internal shocks during the launch of ultra-relativistic jets \citep{1989Natur.340..126E, 1992ApJ...395L..83N, 1997ApJ...490...92K}. They can last from tens of milli-seconds to thousands of seconds, and large samples of GRBs show a bimodality in their duration as a result of different progenitor systems \citep{1993ApJ...413L.101K}. LIGO's GW170817 combined with the Fermi detection of short GRB 170817A confirmed that at least some of the short GRB population are produced by merging neutron stars \citep{2017ApJ...848L..13A}. Long GRBs have often been observed in conjunction with supernovae and so are thought to be produced during the collapse of massive stars \citep{1993ApJ...405..273W, 2003Natur.423..847H}.

The prompt emission, the GRB, is followed by a broadband afterglow component seen from radio wavelengths to TeV energies.  %\citep[e.g.][]{2018ApJ...856L..18M, 2019Natur.575..455M}. 
This emission is interpreted using the \textit{fireball model} \citep{1992MNRAS.258P..41R, 1993ApJ...405..278M, 1999PhR...314..575P}. In the fireball model, the jet (modelled as a blast wave) decelerates as it interacts with the circum-burst medium creating shocks that accelerate electrons into a power law distribution: $N(E)dE \propto E^{-p}dE$, with 2 $<$ p $<$ 3, which produces synchrotron emission as they cool. The synchrotron spectrum comprises of power laws connected by three breaks: the synchrotron self-absorption break $\nu_{SA}$, the frequency emitted by electrons with the lowest energy $\nu_{M}$, and the cooling break $\nu_{C}$ \citep{1998ApJ...497L..17S}. The evolution of the spectrum is dependent on the fraction of the energy going into the electrons and magnetic fields, and the density profile of the surrounding circum-burst medium: whether it is interstellar medium (ISM)-like i.e. constant density or stellar wind-like - it has a density profile ($\rho = A r^{-2}$).

The radio afterglows of long and short GRBs have different luminosity ranges. Short GRBs tend to be fainter, falling below $10^{30}$erg~s\textsuperscript{-1}Hz\textsuperscript{-1} \citet{2020arXiv200808593F}. This restricts the redshift range in which short GRBs are detectable, the most distant radio detected short GRB is GRB 141212A at only z=0.596 \citep{2014GCN.17177....1C}. In total only eight on-axis short GRBs have been detected at radio frequencies, each event was only observable for around 10 days \citep{ 2005Natur.438..988B,2006ApJ...650..261S, 2014ApJ...780..118F, 2015ApJ...815..102F,2016ApJ...829....7L, 2019ApJ...883...48L, 2020arXiv200808593F}. Their afterglows are dominated by the `forward shock' (FS), a shock which propagates out into the surrounding circum-burst medium and is observed at all wavelengths \citep[e.g.][]{2016ApJ...827..102T}. The rarity of radio short GRB detections, hinders our understanding in their evolution as we depend predominantly on X-ray and optical data to study these systems.

Long GRBs are comparatively more luminous $>10^{30}$erg~s\textsuperscript{-1}Hz\textsuperscript{-1}, and therefore seen out to higher redshifts and for longer, some of these events have even been detected on the timescales of years \citep{2012ApJ...746..156C, 2008A&A...480...35V}. In the afterglows of some long GRBs, a second shock is often observed: the `reverse shock', which is mostly observed at optical and sometimes radio wavelengths on the timescale of days \citep[e.g.][]{2014MNRAS.444.3151V, 2019MNRAS.486.2721B}. The reverse shock accelerates electrons as it travels back towards the newly formed compact object.

The expected flux evolution of some afterglow light curves are cut short by a jet break. A jet break is observed when the jet has decelerated enough such that, $\Gamma < 1/\theta_{j}$ where $\Gamma$ is the bulk Lorentz factor of the jet and $\theta_{j}$ is the opening angle of the jet. Around this point, instead of expanding radially, the jet starts expanding laterally. 
Once this transition occurs: for optically thin emission the light curve decays rapidly at $F \propto t^{-p}$ \citep{1999ApJ...519L..17S, 2018ApJ...868L..11M}. For optically thick jets ($\nu < \nu_{SA}$), the light curve flattens, and emission between $\nu_{SA} < \nu < \nu_{M}$ the light curve decays at $F \propto t^{-1/3}$.

GRB 200826A was first detected by the Fermi Gamma-ray Burst Monitor at 04:29:52 UT on 2020 August 26 (T\textsubscript{0}) \citep{2020GCN.28284....1F}. With a T\textsubscript{90}=1.1$\pm$0.1s between 50-300\,keV, it was classified as a short GRB. The \textit{Swift} X-Ray Telescope (XRT) reported seven uncatalogued sources within the Fermi error region \citep{2010A&A...519A.102E}. A potential afterglow candidate, ZTF20abwysqy, was identified at redshift 0.714$\pm$0.137 by the Zwicky Transient Facility \citep[ZTF,][]{2020GCN.28295....1A} with coordinates consistent with source three from \textit{Swift}-XRT. \citet{2020GCN.28302....1A} reported the first radio detection of this source at 2.28\,days post burst with a flux density of $\sim$40\,$\mu$Jy at 6\,GHz. Since the initial detection, further spectral analysis by \citet{2020GCN.28301....1S} showed that GRB 200826A may have been a long GRB at the short end of the T\textsubscript{90} distribution. Optical observations using the Gemini North telescope have detected emission bright enough to originate from a supernova as opposed to a kilonova, the thermal counterpart associated with short GRBs, providing further evidence that this may be a long GRB \citep{2020GCN.28727....1A}.

In this paper, we present our radio observations and discuss their interpretation in a way that is applicable to both long and short GRBs. We use standard $\Lambda$CDM cosmology: $\Omega_{M} = 0.3$, $H_{0} = 70$kms\textsuperscript{-1}Mpc\textsuperscript{-1} and $\Omega_{\Lambda} = 0.7$. 

\section{Observations}

Observations of GRB 200826A with the \textit{enhanced Multi Element Remotely Linked Interferometer Network} (eMERLIN) were obtained through proposal CY10002 (PI: Rhodes). The field of GRB 200826A was observed for six separate epochs between four and eleven days post-burst at 5\,GHz, with a bandwidth of 512\,MHz. All dishes except for the Lovell were used. Each measurement set was averaged down to 4\,second integrations and 512 channels. A priori flags were applied due to RFI followed up with additional flagging to improve the quality of datasets.. Calibration was performed using the eMERLIN pipeline\footnote{https://github.com/e-merlin/eMERLIN\_CASA\_pipeline}. Initial bandpass calibration was performed using J1407+2827, before calculating complex gains using J0012+3353. Absolute flux scaling was applied from 3C 286. Calibration tables were then applied to the target field. The calibrated measurement set was imaged using \textit{tclean} in \textsc{casa} \citep{casa}.

\section{Results}

Here, we present the results of the previously described eMERLIN radio observations, along side publicly available data from the Karl G. Jansky Very Large Array \citep[VLA,][]{2020GCN.28302....1A}, the upgraded Giant Meterwave Radio Telescope (uGMRT), \citep{2020GCN.28410....1C} and the \textit{Swift} X-ray Telescope (XRT). The data are interpreted in the context of the fireball model by fitting power law components to the data. All results are presented following the convention $F_{\nu} \propto t^{\alpha}\nu^{\beta}$, \textit{t} is the time since burst, $\nu$ is the central frequency or energy of the observing band, and $\alpha$ and $\beta$ are the power law indices.
 
\subsection{X-ray}
\label{sec:xobs}

The afterglow candidate of GRB 200826A was observed by \textit{Swift}-XRT from $\sim$0.7 to 8 days post burst \citep{2020GCN.28300....1D} in the 0.3-10\,keV band.  
The XRT data points are shown as black filled circles in the upper panel of Figure \ref{fig:lightcurve}. All data points have been corrected for absorption. We fit the light curve with a single power law decay. The decay follows F$\propto t^{-1.8\pm0.4}$ and is shown in Figure \ref{fig:lightcurve} as the green dot-dashed line. We measure a reduced $\chi$-squared of 3.5. We note that the last data point shows an excess flux with respect to the given model. This lends itself to the possibility that the light curve could also be fit with a broken power law, however an f-test performed on the data set allow us to reject a broken power law in favour of a single power law fit \citep{2009MNRAS.397.1177E}. 

The \textit{Swift} burst analyser fit an absorbed power law spectrum to each GRB 200826A epoch \citep{2010A&A...519A.102E}. The lower panel of Figure \ref{fig:lightcurve} shows the photon index evolution, i.e. the power law fit to each spectrum. There appears to be no significant evolution of photon index over the period where GRB 200826A is detected. The average photon index is $1.5\pm0.2$, corresponding a spectral index of $\beta = -0.5\pm0.2$ (green horizontal dot-dashed line in the lower panel of Figure \ref{fig:lightcurve}).

All flux and photon index results are given in Table \ref{tab:xobs}.

\begin{table}
 
 \begin{center}
 \begin{tabular}{cccc}
  \hline
  \hline
   T-T\textsubscript{0}(days) & $\Delta$T (days) & Flux Density ($\mu$Jy)  & Photon Index\\
  \hline
  \hline
  0.70 & 0.01 & 0.4 $\pm$  0.1 & 1.2 $\pm$ 0.4 \\
  0.75 & 0.01 & 0.035 $\pm$  0.009 & 1.2 $\pm$0.3\\
  0.8 & 0.1 & 0.028 $\pm$ 0.007 & 1.3 $\pm$  0.3 \\
  1.8 & 0.3 & 0.005 $\pm$  0.001 & 1.6 $\pm$  0.3\\
  2.6 & 0.3 & 0.0035 $\pm$  0.0008 & 1.8 $\pm$  0.4\\
  6.4 & 2.2 & 0.0027 $\pm$  0.0006 & 1.7 $\pm$  0.4\\
  \hline
 \end{tabular}
 \label{tab:xobs}
  \caption{\textit{Swift}-XRT flux densities at 10\,keV and photon indices between 0.3-10\,keV. `T' is the time in the middle of each observation and `T\textsubscript{0}' is 04:29:52 UT on 2020 August 26 \citep{2020GCN.28284....1F}. $\Delta$T (days) reflects the duration of each observation \citep{2007A&A...469..379E}. }
 \end{center}
\end{table}

\subsection{Radio}
\label{sec:radioobs}

The first eMERLIN observation of the GRB 200826A field, starting 4.67 days post burst, showed a point source with coordinates: (J2000) R.A. 00$^{\rm h}$27$^{\rm m}$08.54$^{\rm s}$ and Dec +34$^{\circ}$01'38.34". The positional uncertainty is $\pm0.01$". The location of the VLA source reported by \citet{2020GCN.28302....1A} is consistent with our more precisely measured position. 
The remaining five epochs were non-detections, when examined individually. To increase the possibility of detecting the afterglow, epochs two and three were combined into a single longer observation to reduce the rms noise in the field resulting in a 4$\sigma$ detection. The two initial observations had rms noise levels of 19 and 34\,$\mu$Jy/beam, respectively. Epochs four, five and six, which had rms levels of 22, 14 and 16\,$\mu$Jy/beam, were also combined but still showed no detection. The flux densities and non-detections are shown in Table \ref{tab:radioobs}. 

\begin{table}
 \begin{center}
 \begin{tabular}{cccc}
  \hline
  \hline
   T-T\textsubscript{0}(days) & Flux Density ($\mu$Jy) & T\textsubscript{c}-T\textsubscript{0}(days) & Flux Density ($\mu$Jy)  \\
  \hline
  \hline
  4.92$\pm$0.5 & $93\pm16$ & - & - \\
  \hline
  5.91$\pm$0.46 & $<57$ &\multirow{2}{5em}{$\ \ 6.4\pm0.9$\textsuperscript{a}}  &\multirow{2}{4em}{$\ \ 68\pm8$}\\
  6.90$\pm$0.42 &  $<102$\\
  \hline
  7.65$\pm$0.83 & $<66$ & \multirow{3}{5em}{$\ \ 8.7\pm1.9$\textsuperscript{b}} & \multirow{3}{4em}{$\ \ <34$}\\
  8.92$\pm$0.79 & $<42$\\
  9.96$\pm$0.67 & $<48$\\
  \hline
  \label{tab:radioobs}
  %\hline
 \end{tabular}
 \caption{Table of 5\,GHz observations from eMERLIN. `T' is the time in the middle of each observation and `T\textsubscript{0}' is 04:29:52 UT on 2020 August 26 \citep{2020GCN.28284....1F}. In order to reach lower noise levels, we concatenated some of data sets. The central time of each concatenated dataset is shown in the column labelled T\textsubscript{c}. \textsuperscript{a}: a concatenation of epochs two and three starting at 5.91 and 6.90 days post burst. \textsuperscript{b}: epochs four, five and six combined. The errorbars quoted with the observation times reflect the duration of each observation.  Any flux density value prefixed by `$<$' is a 3$\sigma$ upper limit. }
 \end{center}
\end{table}

\begin{figure*}
    \centering
    \includegraphics[width=0.8\textwidth]{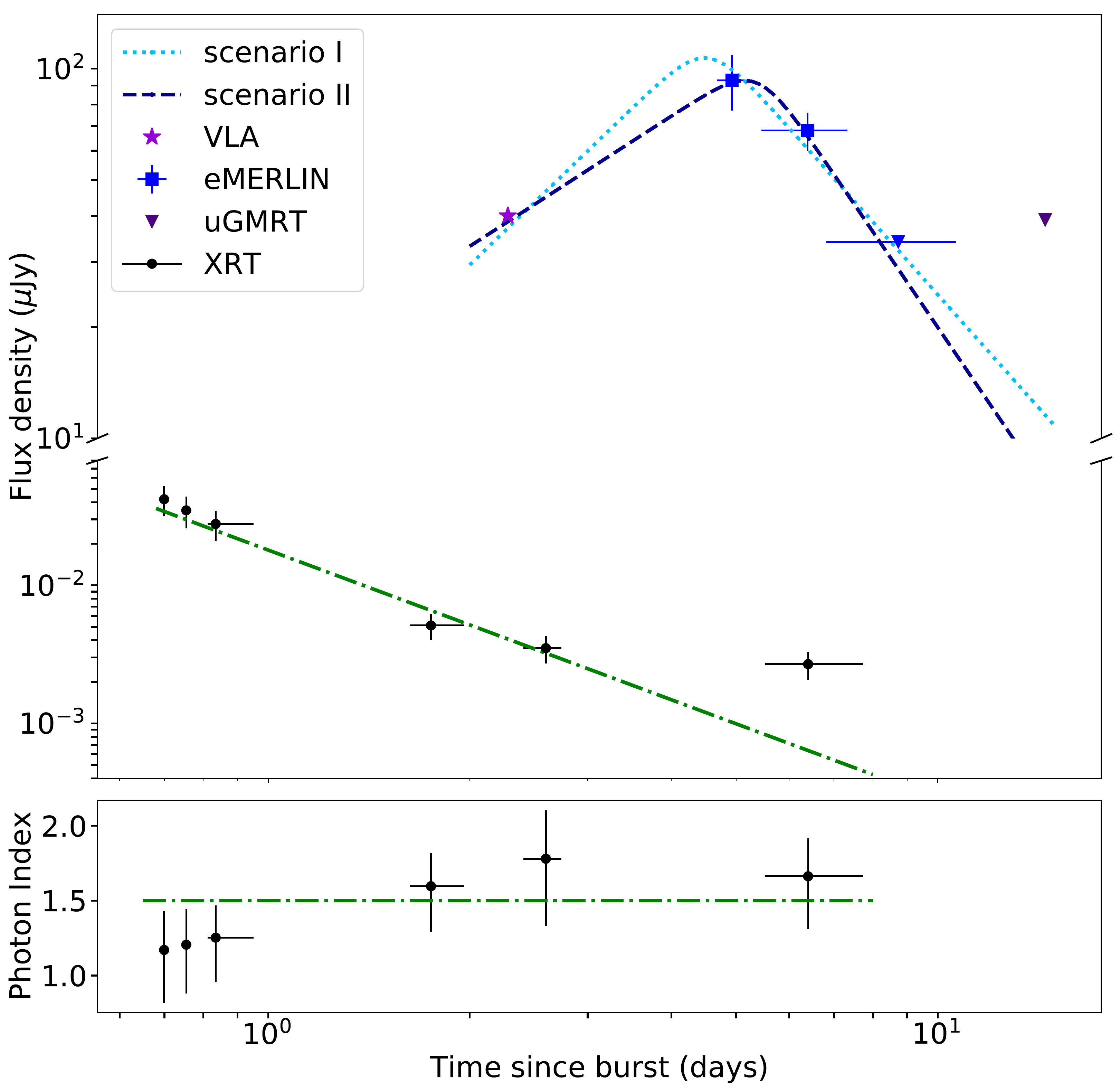}
    \caption{\textit{Upper panel}: Radio light curve of GRB 200829A combining eMERLIN 5GHz data with \citet{2020GCN.28302....1A}'s VLA reported flux and \citet{2020GCN.28410....1C}'s uGMRT upper limit as shown in blue squares and upside down triangle, the light purple star, and dark purple upside down triangle, respectively. The horizontal error bars on the eMERLIN data points show the duration of the observations required to make the respective detections or non-detections. The uGMRT data point is scaled from 1.25\,GHz to 5\,GHz using a spectrum that transitions from optically thick to thin. The light blue dotted and dark blue dashed lines correspond to the models presented in sections \ref{sec:scenario1} and \ref{sec:scenario2}, respectively. The black circles are from \textit{Swift}-XRT showing a power law decay - the green dot-dashed line. \textit{Lower panel}: The spectrum was analysed for each epoch and fit using an absorbed power law with an average photon index: $ 1.5\pm0.2$, denoted by a green dot-dashed line.} 
    \label{fig:lightcurve}
\end{figure*}

The radio light curve in upper panel of Figure \ref{fig:lightcurve}, shows our eMERLIN data combined with that from the VLA, and the uGMRT. The eMERLIN data points are blue squares and the downward facing triangle, \citet{2020GCN.28302....1A}'s VLA detection is shown as the light purple star and the uGMRT upper limit is given as a dark purple downwards-facing triangle \citep{2020GCN.28410....1C}. The VLA observation was made at a central frequency of 6\,GHz but with a wide bandwidth of 4\,GHz, there is comfortably enough overlap between the VLA and eMERLIN observing frequencies. The uGMRT observation was made at 1.25\,GHz and so we have scaled this upper limit of 48.6\,$\mu$Jy to 39\,$\mu$Jy at 5\,GHz. In scaling the upper limit, we have considered the position of synchrotron self-absorption frequency, discussed in section \ref{sec:scenario1}, at 2.1\,GHz at the time of the observation and corrected the flux density accordingly. For the analysis of this data set, given the low flux level of the source we assume a 15\% error on the VLA point.

\section{Discussion}

The XRT light curve, the upper panel of Figure \ref{fig:lightcurve}, may be from synchrotron emission from a FS either above or below the cooling break. The expected temporal exponent for emission from above the cooling break, where synchrotron losses are significant, is $\alpha = \frac{2-3p}{4}$, independent of circum-burst environment, which for a measured $\alpha_{10\textrm{\,keV}} = -1.8\pm0.4$ gives $p = 3.1\pm0.7$. The temporal exponent for emission below the cooling break is $\alpha = \frac{3(1-p)}{4}$ and $\frac{1-3p}{4}$ for an ISM and wind environment, respectively \citep{2002ApJ...568..820G}. Comparing these exponents to that measured would give $p = 3.4\pm0.8$ and $p = 2.7\pm0.6$. Within uncertainties, all the above values of \textit{p} fall within the expected range \citep{2019MNRAS.489.1919T}. 

To determine which branch of the synchrotron SED the emission detected by XRT originates from, we look at the X-ray spectrum. An average photon index of $1.5\pm0.2$, gives $\beta = -0.5\pm0.2$, which is too shallow to originate from above the cooling break. Below the cooling break, this value of $\beta$ gives $p = 2.0\pm0.8$, one that is more in agreement with the stellar-wind scenario than the ISM case from the X-ray light curve. %The errors on $p$ are such that consistent with the values of $p$ derived from the light curve for both a wind and ISM environment. However, $p$ from the photon index is agrees more with that for a wind environment. 
A shallow spectrum is expected from optically thin synchrotron emission, suggesting that the 0.3-10\,keV emission likely originates from below the cooling break i.e. $10\textrm{keV} < \nu_{C}$. Combining this information with that from the light curve, these data show us that the FS is moving through a medium with a wind-like density profile. 

The radio light curve is more complex to interpret. Firstly, we can use the radio luminosity to help determine whether GRB 200826A is a short or long GRB. The luminosity of the first eMERLIN data point is $1.6^{+0.4}_{-0.6}\times10^{30}\textrm{\,erg\,s\textsuperscript{-1}Hz\textsuperscript{-1}}$. 
Long GRBs of such low luminosity have been detected previously, \citep{2012ApJ...746..156C, 2018MNRAS.473.1512A, 2020MNRAS.496.3326R}, but at higher frequencies and only at very low redshift $\la$0.1, far lower than the redshift of GRB 200826A. On the other hand, comparison to the radio luminosities in figure 13 of \citet{2020arXiv200808593F} shows no significant differences between the luminosity of GRB 200826A and the rest of the radio-detected short GRB population. We acknowledge that the radio-detected short GRB population is very small compared to the corresponding long GRB population.

Due to a span of a few days between the VLA reported detection and our first eMERLIN detection, where we have no radio data, we are limited in our knowledge of the location of the radio light curve peak.  
Therefore, we consider two separate scenarios to interpret these radio data based on the time of the light curve peak. In the first scenario, we assume that the peak of the light curve precedes the first eMERLIN observation but after the VLA epoch. In the second scenario, we use the first eMERLIN data point as the light curve peak.
%%%% scenario number 1

\subsection{Scenario I: Radio peak comes from $\nu_{SA}$ moving through the radio band}
\label{sec:scenario1}

To determine the earliest possible peak of the radio light curve, we require the rise from the VLA point to be as steep as possible without being unphysical. The steepest rise, which would produce the earliest light curve break would occur if the FS shock is optically thick. In a wind-like environment the flux evolves quickly as $F \propto t^{1.75}$ \citep{2002ApJ...568..820G}. The break following the rise would be due to $\nu_{SA}$ passing through the band. In an ISM environment, i.e. one with a constant electron number density, the rise would follow $F \propto t^{1.25}$. This shallower rise would cause the break to occur during our first observation. Additionally, our XRT light curves favour a wind environment, which further disfavours this scenario. If the break was due to $\nu_{M}$, the rise would be far shallower than for an optically thick FS: $F\propto t^{1/2}$ or $t^0$ for an ISM or wind density profile, respectively. Either case is too shallow to be consistent with the radio light curve.

Therefore, we fit a broken power law, with the rise fixed as t\textsuperscript{1.75}, to the data using the Monte Carlo Markov Chain (MCMC) sampler \textsc{emcee} \citep{emcee}. The optimum fit was found using maximum likelihood analysis with flat priors on all variables. We used 700 independent walkers, each taking 10000 steps, the first 6000 of which were burnt. The mean, lower and upper uncertainties quoted from the analysis are the 50\textsuperscript{th}, 16\textsuperscript{th} and 84\textsuperscript{th} percentiles of the samples in the marginalised distributions, respectively. The results from fitting a broken power law to the data this way are consistent with those if we fit a single power law to only the eMERLIN data.

The resulting fit is shown as the light blue dotted line in Figure \ref{fig:lightcurve} (the corner plot of the fit is shown in the Appendix, Figure \ref{sec:corner}). The decay seen in the eMERLIN data can be described using a power law decay of F$\propto t^{-2.0^{+0.6}_{-0.8}}$, denoted by the light blue dotted power law in the upper part of Figure \ref{fig:lightcurve}. Comparison with theoretical light curves from \citet{2014MNRAS.444.3151V}, shows that the eMERLIN data are also in agreement with emission from the optically thin branch of the synchrotron spectrum i.e. on the same branch as the XRT light curve. Equating $\alpha_{\textrm{5\,GHz}}$ with the exponents for an ISM and wind environment, gives $p = 3.7\pm1$ and $p = 3.0^{+0.9}_{-1.2}$, respectively. Only the value of $p$ for a stellar wind environment ($p = 3.0$) is in agreement with the results from our XRT data. Both the X-ray and eMERLIN light curves may have been produced by an optically thin FS propagating through a wind-like density profile circum-burst medium. 

From our light curves, we conclude that the eMERLIN and XRT data sets both are produced by optically thin synchrotron. This is confirmed by measuring a radio-X-ray spectral index. We measure a radio-X-ray spectral index of $\beta_{5\textrm{GHz}-10\textrm{keV}} = -0.52\pm0.01$, one that is consistent with optically thin synchrotron, confirming that both the 5\,GHz and 10\,keV data points originate from the same branch of the synchrotron SED. Our value for $\beta_{5\textrm{GHz}-10\textrm{keV}}$ converts to $p = 2.04\pm0.04$, which is shallower than our radio-derived values but is in agreement from that calculated from the XRT photon indices.

Therefore, for the peak in the light curve to occur before the first data point, we need an optically thick rise through a wind-like environment. The eMERLIN data are in the regime such that $\nu_{M} < \nu_{SA} < 5\,\textrm{GHz} < \nu_{C}$.

From inferring that the peak in the light curve is due to the transition from optically thick to optically thin, we are able to place constraints on the emitting region size and the minimum energy present in the jet as the time of the light curve peak. In addition, by assuming a jet geometry we are able to estimate the bulk Lorentz factor of the jet. Using \citet{2013ApJ...772...78B}, we apply equipartition theory, extended for synchrotron sources with a bulk relativistic velocity, to our radio light curve assuming that we did not directly observe the peak of the light curve but instead it occurred about 4.5\,days post-burst, where the light curve reached a peak flux of about 110\,$\mu$Jy. We assume that $\Gamma = 1/\theta_{j}$ because we have no prior knowledge of the jet geometry. At 4.5\,days, the size of the emitting region is $2\times10^{17}$\,cm, with a minimum energy of $3\times10^{47}$\,erg, we note that these values and all those in the following analysis have large uncertainties and are not quantified due to the number of additional assumptions in the underlying model. We estimate $\Gamma$ to be $\sim$5 at the peak of the light curve, i.e the jet is mildly relativistic at the peak of the light curve. The assumption of $\Gamma \sim 1/\theta_{j}$ means we can use the previous statement to predict an opening angle of $\sim11^{\circ}$. We assume a wind environment and are able to calculate $A_{\star}$, which related to the constant, A, from the assumed density profile $\rho = Ar^{-2}$. We measure $A_{\star}$ = 0.4, making $A = 2\times10^{11}$g\,cm\textsuperscript{-1}, where $A = \dot{M}/4\pi V_{W} = 5 \times 10^{11} A_{\star} \textrm{g~cm}^{-1}$; $\dot{M}$ and $V_{W}$ are the mass loss rate and wind velocity of the progenitor star, respectively, \citep{2000ApJ...536..195C}.

The presence of a wind-like circumburst environment, despite the low afterglow luminosity, lends this event to appear more similar to the afterglows of long GRBs where the jet is propagating through the material blown off the star in the late stages of its lifetime \citep{2000ApJ...536..195C}. This is in contention with the duration of the prompt emission leading to the initial interpretation of GRB 200826A as a short GRB, however, as mentioned in the introduction, analysis of the prompt emission has led to the suggestion that this event is a long GRB at the short end of the prompt emission duration distribution \citep{2020GCN.28301....1S}.

%%%% scenario number 2
\subsection{Scenario II: Jet break occurs around the time of the radio light curve peak}
\label{sec:scenario2}

In our second scenario, we consider the possibility that we have observed the peak in the radio light curve, where the break occurred as late as possible, at the time of the first eMERLIN data point. We fit all the radio data with a broken power law in one instance using \textsc{emcee} \citep{emcee}. 

The results of the MCMC fitting show that the rise of the light curve follows $F \propto t^{1.2\pm0.3}$ to a peak of $90\pm10\,\mu$Jy at $5.4^{+0.5}_{-0.6}$\,days followed by a decay following $F \propto t^{-2.7\pm0.9}$. The results are shown as the dark blue dashed line in Figure \ref{fig:lightcurve} and the corresponding corner plot is Figure \ref{fig:corner_brkn_pl}.

The rise of the radio light curve, $F \propto t^{1.2\pm0.3}$, is shallower than in section \ref{sec:scenario1} and more consistent with optically thick FS emission propagating into an ISM environment. In this second scenario, the decay is steeper compared to the first: $F \propto t^{-2.7\pm0.9}$. The uncertainties here are large due to the close proximity in time between the two eMERLIN detections and the break followed by an upper limit. The decay is consistent with the XRT light curve at a 68\% confidence level ($\alpha_{10\textrm{keV}} = -1.8\pm0.4$). When considered without the XRT result, the steeper eMERLIN decay is not compatible with an optically thin decay. Instead, we suggest a different interpretation, one which is not caused by a SSA turnover but by a jet break. The decay following the break is steep enough to be due to a jet break, where the jet begins to spread laterally causing the flux to decay as $F \propto t^{-p}$ \citep{1999ApJ...519L..17S}. However, when combined with the optically thick rise, the post-break decay should plateau for an optically thick jet or follow a shallow decay of $F \propto t^{-1/3}$ when $\nu_{SA} < \nu < \nu_{M}$, and no decay as steeply as observed \citep{1999ApJ...519L..17S}. The observed break may only be possible if the jet becomes optically thin during the peak, as shown in our first interpretation, $\nu_{SA}$ passes through the band during the break time.

This scenario is further complicated by the fact that jet breaks are achromatic and there is no evidence of a break in the X-ray light curve. The final XRT observation starts before the break in the radio light curve and shows an excess in flux with respect to the single power law decay. We speculate that such an excess, if real, could originate from long lived central engine activity and therefore could hide a jet break \citep{2014MNRAS.439.3916M, 2014ApJ...780..118F}. 

The break in the radio light curve can be used to calculate the opening angle of the jet \citep{1999ApJ...519L..17S, 2001ApJ...562L..55F} with the equation:

    $$\theta_{j} = 9.51 t_{\textrm{j,d}}^{3/8} (1 + z)^{-3/8} E_{\textrm{K, ISO,} 52}^{-1/8} n_{0}^{1/8} \textrm{deg}$$

The isotropic equivalent energy of this event E\textsubscript{K, ISO}\,=\,$4.7\times 10^{51}$erg, where the source is at a redshift of 0.714$\pm$0.137 \citep{2020GCN.28301....1S, 2020GCN.28295....1A}. The luminosity of this GRB is consistent with that of a short GRB, however, given the evidence that this may be a long GRB, we used a range of circumburst density values ($n_0$) to calculate the jet opening angles \citep{2020GCN.28727....1A, 2020GCN.28301....1S}. For short GRBs, we assume $\sim0.01$cm\textsuperscript{-3} and for long GRBs, a higher density environment of $n_0\sim1$cm\textsuperscript{-3} \citep{2012ApJ...746..156C, 2015ApJ...815..102F}. From these values, we calculate a $\theta_j = \sim9\degr$ and $\sim16\degr$ for a short and long GRB, respectively. We do not provide an uncertainty for each $\theta_j$ measurement because of the large assumptions made in addition to the numerical uncertainties on each input value.

Jet break detections from previous short GRB systems, gives $\theta_j = 3-8\degr$, however this range increases to larger opening angles once lower limits are considered, \citep{2015ApJ...815..102F}. Long GRBs $\theta_{j}$ measurements have a larger range at $7.4_{-6.6}^{+11}\degr$ \citep{2014ApJ...781....1L, 2016ApJ...818...18G}. Our calculations of $\theta_j$ for long and short GRBs are in agreement with results for both of their respective populations. Our result for $\theta_j$ in section \ref{sec:scenario1} also sits comfortably within the bounds of both populations.

Comparing the two interpretations we have presented show that our second scenario is far more complex than that described in section \ref{sec:scenario1}. Furthermore, the conclusions reached in section \ref{sec:scenario1} are more consistent with other multi-wavelength observations of this source and so we favour our earlier explanation of the data. 

\section{Summary}
Here, we have reported on radio observations of GRB 200826A performed with eMERLIN at 5\,GHz lasting from four to eleven days after the burst. Initially classed as a short GRB, \citet{2020GCN.28295....1A}'s redshift measurement for GRB 200826A's potential host galaxy would make this event the most distance radio detected short GRB to date. Further analysis into the prompt emission hinted that this maybe a long GRB the short end of the T\textsubscript{90} distribution \citep{2020GCN.28301....1S}. We consider two possible interpretations for the analysis of our eMERLIN detections, used together with a single detection reported from the VLA, \citep{2020GCN.28302....1A}. Both include an optically thick rise although with different density profiles. We consider a range of break times, from as early as possible until our first eMERLIN observation. Breaks occurring at different times have different underlying causes: a break before the first observation may be due to synchrotron self-absorption break passing through the band. If the peak of the light curve occurred during the first eMERLIN epoch, the resulting break may be caused by a jet break combined with a transition from an optically thick to optically thin regime around the same time. We calculate jet opening angle values for both long and short GRB environments deriving 16$\degr$ and 9$\degr$, respectively. No evidence of a break is seen in the X-ray light curve, instead we see optically thin synchrotron emission lasting for the first $\sim$6 days, which is confirmed by the photon indices measured from the spectra at each epoch. Given the relative complexity of our second scenario with respect to the first, we favour the earlier interpretation where the break in the radio light curve originates from the synchrotron self-absorption frequency. Our first scenario is also more consistent with the suggestion that this event was a long GRB with prompt emission lasting less than 2\,seconds, as it shows the jet propagating through a wind-like environment.

\section*{Acknowledgements}

The authors would like to thank the referee for their helpful comments. L. Rhodes acknowledges the support given by the Science and Technology Facilities Council through an STFC studentship. This work made use of data supplied by the UK Swift Science Data Centre at the University of Leicester. We acknowledge the Jodrell Bank Centre for Astrophysics. eMERLIN, funded by the STFC, is a National Facility operated by the University of Manchester at Jodrell Bank Observatory.

%%%%%%%%%%%%%%%%%%%%%%%%%%%%%%%%%%%%%%%%%%%%%%%%%%
\section*{Data Availability}

The data presented in this paper are all available in the article.



\bibliographystyle{mnras}
\bibliography{example} % if your bibtex file is called example.bib

%%%%%%%%%%%%%%%%%%%%%%%%%%%%%%%%%%%%%%%%%%%%%%%%%%

%%%%%%%%%%%%%%%%% APPENDICES %%%%%%%%%%%%%%%%%%%%%

\appendix

\section{MCMC Corner Plots}

\begin{figure}
    \centering
    \includegraphics[width=\columnwidth]{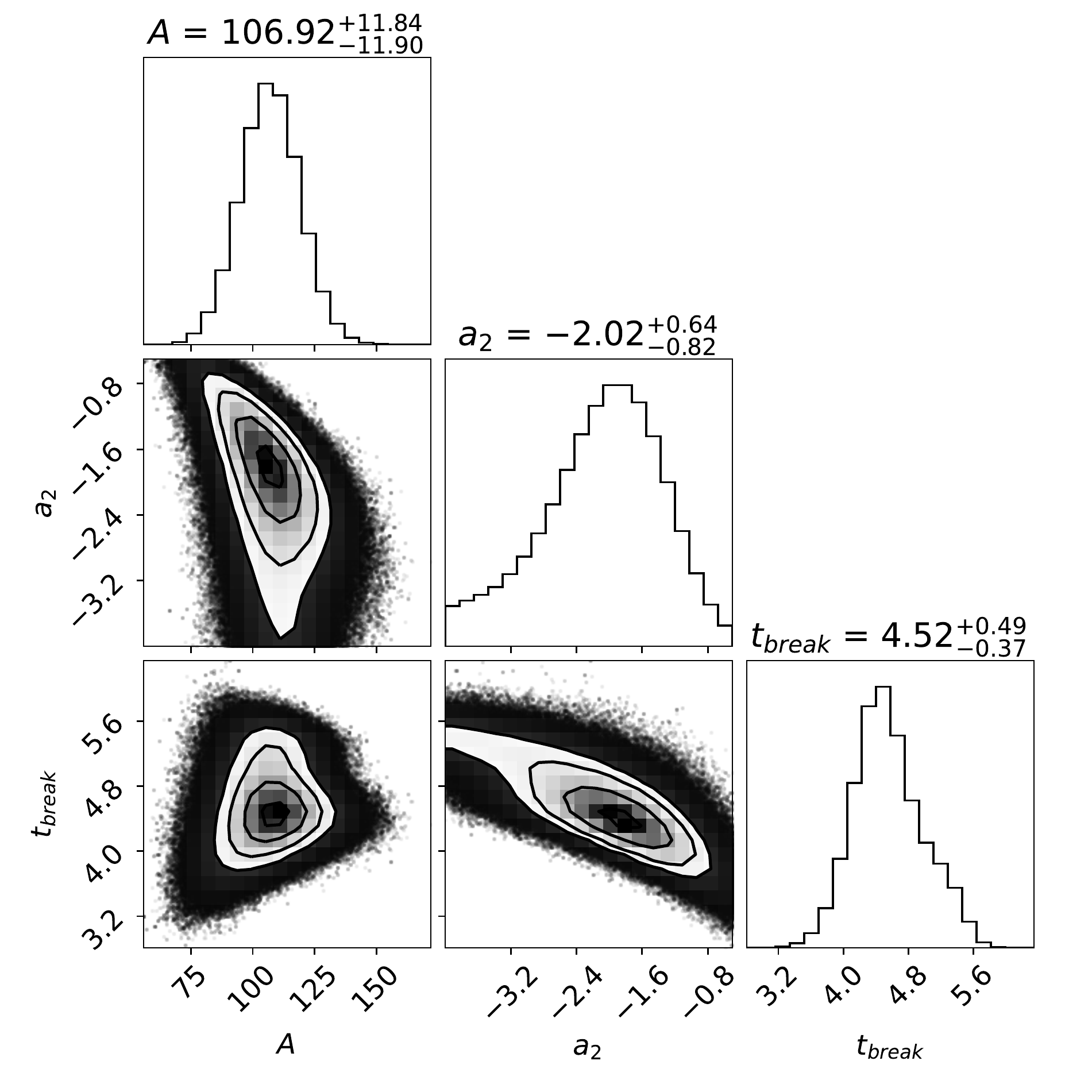}
    \caption{Result of fitting a broken power law with a fixed rise of $t^{1.75}$ using \textsc{emcee} to the eMERLIN light curve.}
    \label{fig:corner_pl}
\end{figure}

\begin{figure}
    \centering
    \includegraphics[width=\columnwidth]{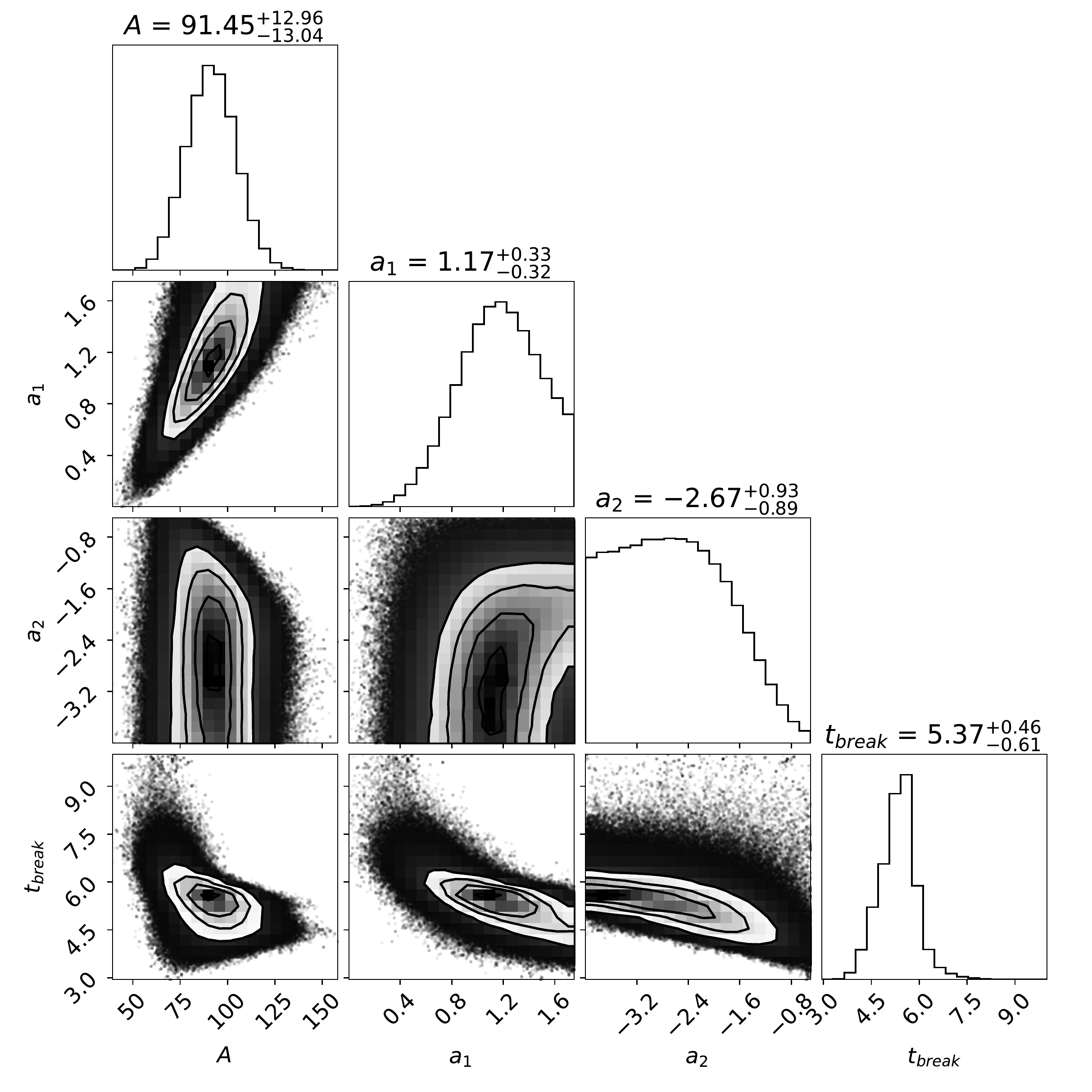}
    \caption{Result of fitting a broken power law to the eMERLIN and VLA data points.}
    \label{fig:corner_brkn_pl}
\end{figure}

The radio data presented in this article are fit with broken power laws. To fit the power laws to the data, we used the Monte Carlo Markov Chain (MCMC) sampler \textsc{emcee} \citep{emcee}. Maximum likelihood analysis was applied to find the optimum fit. Flat priors were used on all variables. We used 700 walkers and burnt the first 6000 of 10000 steps giving us 2.8 million samples. The mean, lower and upper uncertainties quoted from the analysis are the 50\textsuperscript{th}, 16\textsuperscript{th} and 84\textsuperscript{th} percentiles of the samples in the marginalised distributions, respectively. Figures \ref{fig:corner_pl} and \ref{fig:corner_brkn_pl} show the posterior distributions. In Figures \ref{fig:corner_pl} and \ref{fig:corner_brkn_pl}, A is the flux density at the peak of the light curve, t\textsubscript{break} is the time of the light curve break, and a\textsubscript{1} and a\textsubscript{2} are the exponents of the rise and decay power laws, respectively.
\label{sec:corner}

%%%%%%%%%%%%%%%%%%%%%%%%%%%%%%%%%%%%%%%%%%%%%%%%%%

% Don't change these lines
\bsp	% typesetting comment
\label{lastpage}
\end{document}